\documentclass[pdftex,10pt,conference]{IEEEtran}

\usepackage[cmex10]{amsmath} 

\usepackage{latexsym,amssymb,bm,cite,ifthen}
\usepackage[pdftex]{graphicx}

\addtocounter{dbltopnumber}{1}
\addtocounter{totalnumber}{1}

\newcommand{\iftwocolumn}[2]{\ifthenelse{\boolean{@twocolumn}}{#1}{#2}}

\newcommand{\ignore}[1]{}

\newcommand{\Fig}[1]{Fig.~\ref{#1}}

\newcommand{\eqdef}{\stackrel{\scriptscriptstyle\bigtriangleup}{=} }

\newcommand{\cond}{\hspace{0.02em}|\hspace{0.08em}}

\newcommand{\T}{\mathsf{T}}
\renewcommand{\H}{\mathsf{H}}

\newcommand{\C}{{\mathbb C}}

\newcommand{\calY}{\mathcal{Y}}

\newcommand{\ccj}[1]{\overline{#1}}

\newcommand{\tr}{\operatorname{tr}}

\newcounter{examplecntr}
{\begin{trivlist}\small\item[]\refstepcounter{examplecntr}%
 {\bfseries Example~\theexamplecntr%
  \ifthenelse{\equal{#1}{}}{}{ (#1)}.
}}%
{\end{trivlist}}

\newcounter{definitioncntr}
{\begin{trivlist}\item[]\refstepcounter{definitioncntr}%
{\bfseries Definition~\thedefinitioncntr.}}%
{\hfill$\Box$\end{trivlist}}

\newcounter{theoremcntr}
{\begin{trivlist}\item[]\refstepcounter{theoremcntr}%
{\bfseries Theorem~\thetheoremcntr%
  \ifthenelse{\equal{#1}{}}{}{ (#1)}.
}}%
{\hfill$\Box$\end{trivlist}}

\newcounter{propositioncntr}
\newenvironment{proposition}[1][]%
{\begin{trivlist}\item[]\refstepcounter{propositioncntr}%
{\bfseries Proposition~\thepropositioncntr%
  \ifthenelse{\equal{#1}{}}{}{ (#1)}.
}}%
{\hfill$\Box$\end{trivlist}}

\newcommand{\eproofnegspace}{\\[-1.5\baselineskip]\rule{0em}{0ex}}

\newcommand{\cent}[1]{\makebox(0,0){#1}}
\newcommand{\pos}[2]{\makebox(0,0)[#1]{#2}}

\newcommand{\markerDot}{\circle*{1}}

\newcommand{\knownBox}{\cent{\rule{1.75\unitlength}{1.75\unitlength}}}

\begin{document}

\title{A Factor-Graph Representation of\\ Probabilities in Quantum Mechanics%
}

\author{%
\IEEEauthorblockN{Hans-Andrea Loeliger}
\IEEEauthorblockA{ETH Zurich}
\IEEEauthorblockA{loeliger@isi.ee.ethz.ch}
\and
\IEEEauthorblockN{Pascal O.\ Vontobel}
\IEEEauthorblockA{Hewlett--Packard Laboratories, Palo Alto}
\IEEEauthorblockA{pascal.vontobel@ieee.org}
}

\maketitle

\begin{abstract}
A factor-graph representation of quantum-mechani\-cal probabilities 
is proposed. 
Unlike standard statistical models, 
the proposed representation uses auxiliary variables (state variables)
that are not random variables.
\end{abstract}

\section{Introduction}
\label{sec:Intro}

Statistical models with many variables are often represented 
by factor graphs \cite{KFL:fg2000,Lg:ifg2004,LDHKLK:fgsp2007,MeMo:ipc}
or similar graphical representations 
\cite{Jo:gm2004,bi:prml,KoFr:PGMb}. 
Such graphical representations can be helpful in various ways, 
including elucidation of the model itself as well as the derivation
of algorithms for statistical inference. 

So far, however, quantum mechanics (e.g., \cite{NiChuang:QCI,AFP:QM})
has been standing apart. 
Despite being a statistical theory, quantum mechanics does not seem to fit into 
standard statistical categories. 
Indeed, it has often been emphasized that quantum mechanics is a generalization 
of probability theory that cannot be understood in terms of ``classical'' 
statistical modeling.

In this paper, we propose the different perspective 
that the probabilities in quantum mechanics are quite ordinary,
but their state-space representation 
is of a type not previously used in statistical modeling. 
In particular, 
we propose a factor-graph representation of quantum mechanics 
that correctly represents the joint probability distribution
of any number of measurements. 
Like most statistical models, the proposed factor graphs 
use auxiliary variables (state variables) 
in addition to the actually observed variables; 
however, in contrast to standard statistical models, 
the auxiliary variables in the proposed factor graphs are not random variables. 
Nonetheless, the probabilities of the observations 
are marginals of the factor graph, as in standard statistical models.

The paper is structured as follows. 
Section~\ref{sec:FG} reviews factor graphs and their connection to linear algebra and tensor diagrams. 
Section~\ref{sec:StatModel} makes the pivotal observation that factor graphs with complex factors 
and with auxiliary variables that are not random variables
can represent probability mass functions. 
The main results
are given in Section~\ref{sec:QMFG}.

We will use standard linear algebra notation rather 
than the bra-ket notation of quantum mechanics. 
The Hermitian transpose of a complex matrix $A$ 
will be denoted by $A^\H \eqdef \ccj{A^\T}$,
where $A^\T$ is the transpose of $A$ and $\ccj{A}$ is the componentwise complex conjugate. 
An identity matrix will be denoted by $I$.

\section{On Factor Graphs and Matrices}
\label{sec:FG}

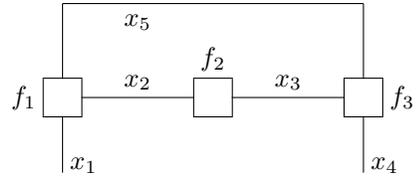
\begin{figure}
\begin{center}
\begin{picture}(55,22.5)(3,7.5)
\put(10,15){\framebox(5,5){}}     \put(9,17.5){\pos{cr}{$f_1$}}
 \put(12.5,15){\line(0,-1){7.5}}    \put(13.5,7.5){\pos{bl}{$x_1$}}
\put(15,17.5){\line(1,0){15}}     \put(22.5,18.5){\pos{cb}{$x_2$}}
\put(30,15){\framebox(5,5){}}     \put(32.5,21){\pos{cb}{$f_2$}}
\put(35,17.5){\line(1,0){15}}     \put(42.5,18.5){\pos{cb}{$x_3$}}
\put(50,15){\framebox(5,5){}}     \put(56,17.5){\pos{cl}{$f_3$}}
 \put(52.5,15){\line(0,-1){7.5}}  \put(53.5,7.5){\pos{bl}{$x_4$}}
\put(12.5,20){\line(0,1){10}}
\put(12.5,30){\line(1,0){40}}     \put(22.5,28.5){\pos{ct}{$x_5$}}
\put(52.5,20){\line(0,1){10}}
\end{picture}
\caption{\label{fig:FactorGraph}%
Forney factor graph of (\ref{eqn:ExampleFactorgraph}).}
\end{center}
\end{figure}

Factor graphs represent factorizations of functions of several variables. 
In this paper, all variables take values in finite alphabets
and all functions take values in $\C$.
We will use Forney factor graphs (also called normal factor graphs) as in \cite{Lg:ifg2004,LDHKLK:fgsp2007}
where nodes/boxes represent factors and edges represent variables.
For example, assume that some function $f(x_1,\ldots,x_5)$
can be written as
\begin{equation}  \label{eqn:ExampleFactorgraph}
f(x_1,\ldots,x_5) = f_1(x_1,x_2,x_5) f_2(x_2,x_3) f_3(x_3,x_4,x_5).
\end{equation}
The corresponding factor graph is shown in \Fig{fig:FactorGraph}.

\begin{figure}
\vskip-0.20cm
\begin{center}
\begin{picture}(55,35)(3,0)
\put(10,15){\framebox(5,5){}}     \put(9,17.5){\pos{cr}{$f_1$}}
 \put(12.5,15){\line(0,-1){15}}    \put(13.5,0){\pos{bl}{$x_1$}}
\put(15,17.5){\line(1,0){15}}     \put(21,18.5){\pos{cb}{$x_2$}}
\put(30,15){\framebox(5,5){}}     \put(32.5,21){\pos{cb}{$f_2$}}
\put(35,17.5){\line(1,0){15}}     \put(42.5,18.5){\pos{cb}{$x_3$}}
\put(50,15){\framebox(5,5){}}     \put(56,17.5){\pos{cl}{$f_3$}}
 \put(52.5,15){\line(0,-1){15}}  \put(53.5,0){\pos{bl}{$x_4$}}
\put(12.5,20){\line(0,1){10}}
\put(12.5,30){\line(1,0){40}}    \put(21,28.5){\pos{ct}{$x_5$}}
\put(52.5,20){\line(0,1){10}}
\put(26,11){\dashbox(36,15){}}     \put(61,27){\pos{bc}{$g$}}  
\put(2,7.5){\dashbox(65,27.5){}}      
\end{picture}
\caption{\label{fig:ClosingBoxes}%
Closing boxes in factor graphs.}
\end{center}
\vskip-0.50cm
\end{figure}
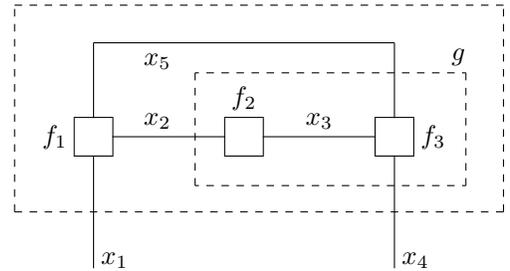

The Forney factor-graph notation is intimately connected 
with the idea of ``closing boxes'' by summing over internal variables \cite{Lg:ifg2004}. 
For example, closing the inner dashed box in \Fig{fig:ClosingBoxes}
replaces the two nodes/factors 
$f_2(x_2,x_3)$ and $f_3(x_3,x_4,x_5)$ by 
the single node/factor
\begin{equation}
g(x_2,x_4,x_5) \eqdef \sum_{x_3} f_2(x_2,x_3) f_3(x_3,x_4,x_5);
\IEEEeqnarraynumspace
\end{equation}
closing the outer dashed box in \Fig{fig:ClosingBoxes}
replaces all nodes/ factors in (\ref{eqn:ExampleFactorgraph})
by the single node/factor
\begin{equation} \label{eqn:FactorGraphExampleMarginal}
f(x_1,x_4) \eqdef \sum_{x_2,x_3,x_5} f(x_1,\ldots,x_5);
\end{equation}
and closing first the inner dashed box and then the outer dashed box
replaces all nodes/factors in (\ref{eqn:ExampleFactorgraph}) by
\begin{equation} \label{eqn:FactorGraphExampleHierachicalMarginal}
\sum_{x_2,x_5} f_1(x_1,x_2,x_5) g(x_2,x_4,x_5)
= f(x_1,x_4).
\end{equation}
Note the equality between (\ref{eqn:FactorGraphExampleHierachicalMarginal}) 
and (\ref{eqn:FactorGraphExampleMarginal}), 
which holds in general: 
closing an inner box within some outer box (by summing over its internal variables) 
does not change the closed-box function of the outer box.

A \emph{half edge} in a factor graph is an edge that is connected to only one node 
(such as $x_1$ in \Fig{fig:FactorGraph}). 
The \emph{external function} 
of a factor graph 
(in \cite{BaMa:nfght2011,FoVo:pfnfg2011c,BMV:nfgla2011c} also called \emph{partition function})
is defined to be 
the closed-box function of a box that contains all nodes and all full edges, 
but all half edges stick out
(such as the outer box in \Fig{fig:ClosingBoxes}). 
The external function of \Fig{fig:FactorGraph} is~(\ref{eqn:FactorGraphExampleMarginal}).

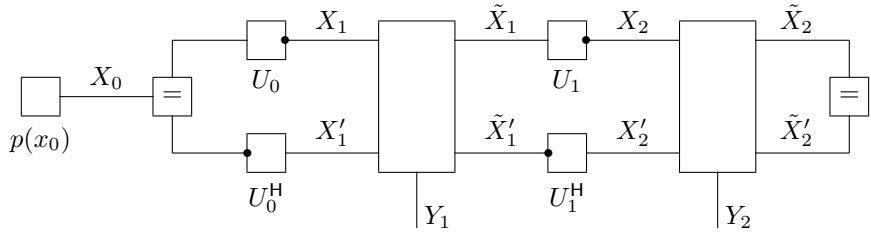
\begin{figure*}[p]
\centering
\begin{picture}(112.5,30)(0,-7.5)
\put(0,7.5){\framebox(5,5){}}      \put(2.5,6){\pos{ct}{$p(x_0)$}}
\put(5,10){\line(1,0){12.5}}       \put(11.25,11){\pos{cb}{$X_0$}}
\put(20,17.5){\line(1,0){10}}
 \put(20,17.5){\line(0,-1){5}}
\put(17.5,7.5){\framebox(5,5){$=$}}
 \put(20,7.5){\line(0,-1){5}}
\put(20,2.5){\line(1,0){10}}
\put(30,15){\framebox(5,5){}}     \put(32.5,13.5){\pos{ct}{$U_0$}}
 \put(35,17.5){\markerDot}
\put(35,17.5){\line(1,0){12.5}}   \put(41.25,18.5){\pos{cb}{$X_1$}}
\put(30,0){\framebox(5,5){}}      \put(32.5,-1.5){\pos{ct}{$U_0^\H$}}
 \put(30,2.5){\markerDot}
\put(35,2.5){\line(1,0){12.5}}    \put(41.25,3.5){\pos{cb}{$X_1'$}}
\put(47.5,0){\framebox(10,20){}}
\put(52.5,0){\line(0,-1){7.5}}    \put(53.5,-7.5){\pos{bl}{$Y_1$}}
\put(57.5,17.5){\line(1,0){12.5}} \put(63.75,18.5){\pos{cb}{$\tilde{X}_1$}}
\put(70,15){\framebox(5,5){}}     \put(72.5,13.5){\pos{ct}{$U_1$}}
 \put(75,17.5){\markerDot}
\put(75,17.5){\line(1,0){12.5}}   \put(81.25,18.5){\pos{cb}{$X_2$}}
\put(57.5,2.5){\line(1,0){12.5}}  \put(63.75,3.5){\pos{cb}{$\tilde{X}_1'$}}
\put(70,0){\framebox(5,5){}}      \put(72.5,-1.5){\pos{ct}{$U_1^\H$}}
 \put(70,2.5){\markerDot}
\put(75,2.5){\line(1,0){12.5}}    \put(81.25,3.5){\pos{cb}{$X_2'$}}
\put(87.5,0){\framebox(10,20){}}
\put(92.5,0){\line(0,-1){7.5}}    \put(93.5,-7.5){\pos{bl}{$Y_2$}}
\put(97.5,17.5){\line(1,0){12.5}}  \put(103,18.5){\pos{cb}{$\tilde{X}_2$}}
\put(110,17.5){\line(0,-1){5}}
\put(107.5,7.5){\framebox(5,5){$=$}}
\put(110,7.5){\line(0,-1){5}}
\put(97.5,2.5){\line(1,0){12.5}}   \put(103,3.5){\pos{cb}{$\tilde{X}_2'$}}
\end{picture}
\caption{\label{fig:GenQFG}%
Factor graph of quantum system with two measurements and the corresponding observations $Y_1$ and $Y_2$.
}
\end{figure*}

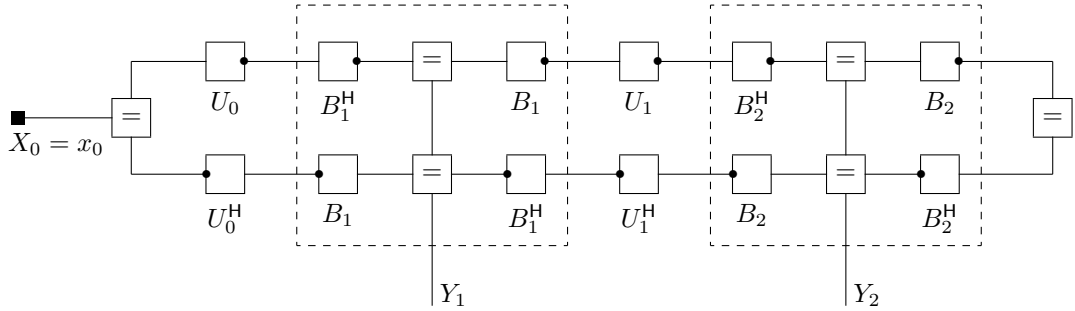
\begin{figure*}
\centering
\begin{picture}(140,40)(-10,-15)
\put(-10,10){\knownBox}
\put(-10,10){\line(1,0){12.5}}
 \put(-5,8){\pos{tc}{$X_0=x_0$}}
\put(5,17.5){\line(1,0){10}}
 \put(5,17.5){\line(0,-1){5}}
\put(2.5,7.5){\framebox(5,5){$=$}}
 \put(5,7.5){\line(0,-1){5}}
\put(5,2.5){\line(1,0){10}}
\put(15,15){\framebox(5,5){}}        \put(17.5,13.5){\pos{ct}{$U_0$}}
 \put(20,17.5){\markerDot}
\put(20,17.5){\line(1,0){10}}
\put(30,15){\framebox(5,5){}}        \put(32.5,13.5){\pos{ct}{$B_1^\H$}}
 \put(35,17.5){\markerDot}
\put(35,17.5){\line(1,0){7.5}}
\put(42.5,15){\framebox(5,5){$=$}}
 \put(45,15){\line(0,-1){10}}
\put(47.5,17.5){\line(1,0){7.5}}
\put(55,15){\framebox(5,5){}}        \put(57.5,13.5){\pos{ct}{$B_1$}}
 \put(60,17.5){\markerDot}
\put(60,17.5){\line(1,0){10}}
\put(70,15){\framebox(5,5){}}        \put(72.5,13.5){\pos{ct}{$U_1$}}
 \put(75,17.5){\markerDot}
\put(75,17.5){\line(1,0){10}}
\put(85,15){\framebox(5,5){}}        \put(87.5,13.5){\pos{ct}{$B_2^\H$}}
 \put(90,17.5){\markerDot}
\put(90,17.5){\line(1,0){7.5}}
\put(97.5,15){\framebox(5,5){$=$}}
 \put(100,15){\line(0,-1){10}}
\put(102.5,17.5){\line(1,0){7.5}}
\put(110,15){\framebox(5,5){}}       \put(112.5,13.5){\pos{ct}{$B_2$}}
 \put(115,17.5){\markerDot}
\put(115,17.5){\line(1,0){12.5}}
\put(127.5,17.5){\line(0,-1){5}}
\put(125,7.5){\framebox(5,5){$=$}}
\put(127.5,7.5){\line(0,-1){5}}
\put(115,2.5){\line(1,0){12.5}}
\put(27,-7){\dashbox(36,32){}}
\put(82,-7){\dashbox(36,32){}}
\put(15,0){\framebox(5,5){}}         \put(17.5,-1.5){\pos{ct}{$U_0^\H$}}
 \put(15,2.5){\markerDot}
\put(20,2.5){\line(1,0){10}}
\put(30,0){\framebox(5,5){}}         \put(32.5,-1.5){\pos{ct}{$B_1$}}
 \put(30,2.5){\markerDot}
\put(35,2.5){\line(1,0){7.5}}
\put(42.5,0){\framebox(5,5){$=$}}
 \put(45,0){\line(0,-1){15}}         \put(46,-15){\pos{bl}{$Y_1$}}
\put(47.5,2.5){\line(1,0){7.5}}
\put(55,0){\framebox(5,5){}}         \put(57.5,-1.5){\pos{ct}{$B_1^\H$}}
 \put(55,2.5){\markerDot}
\put(60,2.5){\line(1,0){10}}
\put(70,0){\framebox(5,5){}}         \put(72.5,-1.5){\pos{ct}{$U_1^\H$}}
 \put(70,2.5){\markerDot}
\put(75,2.5){\line(1,0){10}}
\put(85,0){\framebox(5,5){}}         \put(87.5,-1.5){\pos{ct}{$B_2$}}
 \put(85,2.5){\markerDot}
\put(90,2.5){\line(1,0){7.5}}
\put(97.5,0){\framebox(5,5){$=$}}
 \put(100,0){\line(0,-1){15}}        \put(101,-15){\pos{bl}{$Y_2$}}
\put(102.5,2.5){\line(1,0){7.5}}
\put(110,0){\framebox(5,5){}}        \put(112.5,-1.5){\pos{ct}{$B_2^\H$}}
 \put(110,2.5){\markerDot}
\end{picture}
\caption{\label{fig:BasicQMFG}%
Important special case of \Fig{fig:GenQFG}:
all matrices are unitary and the initial state $X_0=x_0$ is known. 
In quantum-mechanical terms, 
such measurements are projection measurements with one-dimensional eigenspaces. 
}
\end{figure*}

\begin{figure*}[p]
\centering
\begin{picture}(120,40)(-5,-10)
\put(0,7.5){\framebox(5,5){}}      \put(2.5,6){\pos{ct}{$p(x_0)$}}
\put(5,10){\line(1,0){12.5}}       \put(11.25,11){\pos{cb}{$X_0$}}
\put(20,17.5){\line(1,0){10}}
 \put(20,17.5){\line(0,-1){5}}
\put(17.5,7.5){\framebox(5,5){$=$}}
 \put(20,7.5){\line(0,-1){5}}
\put(20,2.5){\line(1,0){10}}
\put(30,15){\framebox(5,5){}}     \put(32.5,13.5){\pos{ct}{$U_0$}}
 \put(35,17.5){\markerDot}
\put(35,17.5){\line(1,0){12.5}}   \put(42,18.5){\pos{cb}{$X_1$}}
\put(30,0){\framebox(5,5){}}      \put(32.5,-1.5){\pos{ct}{$U_0^\H$}}
 \put(30,2.5){\markerDot}
\put(35,2.5){\line(1,0){12.5}}    \put(42,3.5){\pos{cb}{$X_1'$}}
\put(47.5,0){\framebox(10,20){}}
\put(52.5,0){\line(0,-1){7.5}}    \put(53.5,-7.5){\pos{bl}{$Y_1$}}
\put(57.5,17.5){\line(1,0){12.5}} \put(63,18.5){\pos{cb}{$\tilde{X}_1$}}
\put(70,15){\framebox(5,5){}}     \put(72.5,13.5){\pos{ct}{$U_1$}}
 \put(75,17.5){\markerDot}
\put(75,17.5){\line(1,0){12.5}}   \put(81.25,18.5){\pos{cb}{$X_2$}}
\put(57.5,2.5){\line(1,0){12.5}}  \put(63,3.5){\pos{cb}{$\tilde{X}_1'$}}
\put(70,0){\framebox(5,5){}}      \put(72.5,-1.5){\pos{ct}{$U_1^\H$}}
 \put(70,2.5){\markerDot}
\put(75,2.5){\line(1,0){12.5}}    \put(81.25,3.5){\pos{cb}{$X_2'$}}
\put(87.5,0){\framebox(10,20){}}
\put(92.5,0){\line(0,-1){7.5}}    \put(93.5,-7.5){\pos{bl}{$Y_2$}}
\put(97.5,17.5){\line(1,0){12.5}}  \put(103,18.5){\pos{cb}{$\tilde{X}_2$}}
\put(110,17.5){\line(0,-1){5}}
\put(107.5,7.5){\framebox(5,5){$=$}}
\put(110,7.5){\line(0,-1){5}}
\put(97.5,2.5){\line(1,0){12.5}}   \put(103,3.5){\pos{cb}{$\tilde{X}_2'$}}
\put(-5,-10){\dashbox(43,35){}}    \put(16.5,26.5){\pos{cb}{$\rho_1$}}
\put(67,-10){\dashbox(48,35){}}    \put(91,26.5){\pos{cb}{$f_=$}}
\end{picture}
\caption{\label{fig:BackwardMesg}%
The closed-box function of the dashed box on the left is the density matrix $\rho_1(x_1,x_1')$.
If $Y_2$ is not known, the dashed box on the right reduces to the constraint $\tilde{X}_1=\tilde{X}_1'$.
}
\end{figure*}
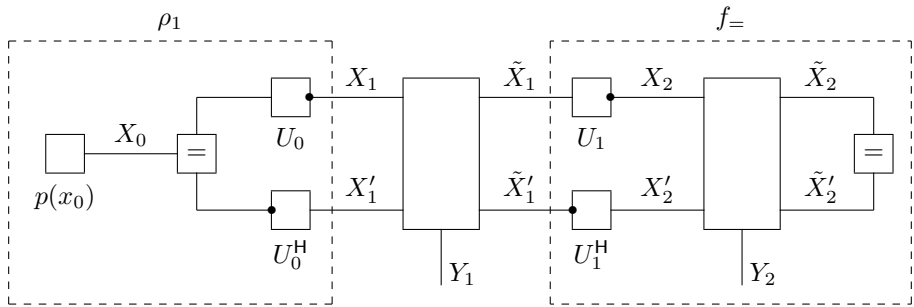

The equality constraint function $f_=$ is defined as
\begin{equation} \label{eqn:EqualityConstraint}
f_=(x_1,\ldots,x_n) = \left\{
 \begin{array}{ll}
   1, & \text{if $x_1= \cdots = x_n$} \\
   0, & \text{otherwise.}
 \end{array}
\right.
\end{equation}
The corresponding node (which is denoted by ``$=$'') can serve
as a branching point in a factor graph, cf.\ Figs.\ \ref{fig:GenQFG}--\ref{fig:GenMeasurement}.

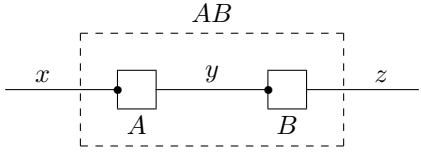
\begin{figure}
\begin{center}
\begin{picture}(55,19)(0,0)
\put(0,7.5){\line(1,0){15}}     \put(5,8.5){\pos{cb}{$x$}}
\put(15,5){\framebox(5,5){}}    \put(17.5,4){\pos{ct}{$A$}}
 \put(15,7.5){\markerDot}
\put(20,7.5){\line(1,0){15}}    \put(27.5,8.5){\pos{cb}{$y$}}
\put(35,5){\framebox(5,5){}}    \put(37.5,4){\pos{ct}{$B$}}
 \put(35,7.5){\markerDot}
\put(40,7.5){\line(1,0){15}}    \put(50,8.5){\pos{cb}{$z$}}
\put(10,0){\dashbox(35,15){}}   \put(27.5,16.5){\pos{cb}{$AB$}}

\end{picture}
\caption{\label{fig:FactorGraphMatrixMult}%
Factor-graph representation of matrix multiplication~(\ref{eqn:MatrixMult}).
The small dot denotes the variable that indexes the rows of the corresponding matrix.
}
\end{center}
\end{figure}

A matrix $A\in \C^{m\times n}$ may be viewed as a function 
\begin{equation}
\{ 1,\ldots, m\} \times \{ 1,\ldots, n\} \rightarrow \C:\, (x,y) \mapsto A(x,y).
\end{equation}
The multiplication of two matrices $A$ and $B$ 
can then be written as
\begin{equation} \label{eqn:MatrixMult}
(AB)(x,z) = \sum_{y} A(x,y) B(y,z),
\end{equation}
which is the closed-box function (the external function) of \Fig{fig:FactorGraphMatrixMult}.
Note that the identity matrix corresponds to an equality constraint 
function $f_=(x,y)$.

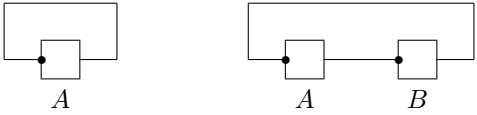
\begin{figure}
\begin{center}
\begin{picture}(15,14)(0,-4)
\put(0,10){\line(1,0){15}}
\put(0,10){\line(0,-1){7.5}}
\put(0,2.5){\line(1,0){5}}
\put(5,0){\framebox(5,5){}}     \put(7.5,-4){\pos{cb}{$A$}}
 \put(5,2.5){\markerDot}
\put(10,2.5){\line(1,0){5}}
\put(15,10){\line(0,-1){7.5}}
\end{picture}
\hspace{15mm}
\begin{picture}(30,14)(0,-4)
\put(0,10){\line(1,0){30}}
\put(0,10){\line(0,-1){7.5}}
\put(0,2.5){\line(1,0){5}}
\put(5,0){\framebox(5,5){}}     \put(7.5,-4){\pos{cb}{$A$}}
 \put(5,2.5){\markerDot}
\put(10,2.5){\line(1,0){10}}
\put(20,0){\framebox(5,5){}}     \put(22.5,-4){\pos{cb}{$B$}}
 \put(20,2.5){\markerDot}
\put(25,2.5){\line(1,0){5}}
\put(30,10){\line(0,-1){7.5}}
\end{picture}
\caption{\label{fig:FactorGraphTrace}%
Factor graph of $\tr(A)$ (left) and of $\tr(AB) = \tr(BA)$ (right).}
\end{center}
\end{figure}

In this notation, 
the trace of a square matrix $A$ is 
\begin{equation} \label{eqn:Trace}
\tr(A) = \sum_{x} A(x,x),
\end{equation}
which is the external function (which is a constant)
of the factor graph in \Fig{fig:FactorGraphTrace}. 
Also shown in \Fig{fig:FactorGraphTrace}
is the graphical proof of the identity $\tr(AB) = \tr(BA)$,
which is much used in quantum mechanics.

Factor graphs for linear algebra operations 
such as \Fig{fig:FactorGraphMatrixMult}
and \Fig{fig:FactorGraphTrace} (and the corresponding generalizations to tensors) 
are essentially tensor diagrams (or trace diagrams) as in \cite{Cv:gtbt,Pe:ula2009}. 
This connection between factor graphs and tensor diagrams 
was noted in \cite{BaMa:nfght2011,FoVo:pfnfg2011c,BMV:nfgla2011c}.

\section{Statistical Models with Auxiliary Variables Using Complex Factors}
\label{sec:StatModel}

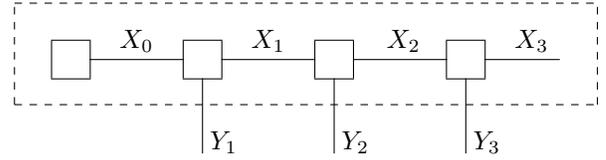
\begin{figure}
\begin{center}
\begin{picture}(67.5,20)(0,0)
\put(0,10){\framebox(5,5){}}
\put(5,12.5){\line(1,0){12.5}}     \put(11.25,13.5){\pos{cb}{$X_0$}}
\put(17.5,10){\framebox(5,5){}}
 \put(20,10){\line(0,-1){10}}      \put(21,0){\pos{bl}{$Y_1$}}
\put(22.5,12.5){\line(1,0){12.5}}  \put(28.75,13.5){\pos{cb}{$X_1$}}
\put(35,10){\framebox(5,5){}}
 \put(37.5,10){\line(0,-1){10}}   \put(38.5,0){\pos{bl}{$Y_2$}}
\put(40,12.5){\line(1,0){12.5}}    \put(46.75,13.5){\pos{cb}{$X_2$}}
\put(52.5,10){\framebox(5,5){}}
 \put(55,10){\line(0,-1){10}}     \put(56,0){\pos{bl}{$Y_3$}}
\put(57.5,12.5){\line(1,0){10}}    \put(63.75,13.5){\pos{cb}{$X_3$}}
\put(-5,6.5){\dashbox(77.5,13.5){}}
\end{picture}
\caption{\label{fig:HMM}%
Factor graph of the hidden Markov model~(\ref{eqn:HMM}) for $n=3$.}
\end{center}
\end{figure}

Statistical models usually contain auxiliary variables 
in addition to the observable variables. 
Consider, for example, a hidden Markov model 
with observables $Y_1,\ldots,Y_n$ and auxiliary variables (hidden variables) $X_0,X_1,\ldots, X_n$ 
such that
\begin{equation} \label{eqn:HMM}
p(y_1,\ldots,y_n,x_0,\ldots,x_n) = p(x_0) \prod_{k=1}^n p(y_k,x_k\cond x_{k-1}). 
\end{equation}
The factor graph of (\ref{eqn:HMM}) is given in \Fig{fig:HMM}. 
(As shown in this example,
variables in factor graphs are often denoted by capital letters \cite{Lg:ifg2004}.)
Closing the dashed box in \Fig{fig:HMM} yields $p(y_1,\ldots,y_n)$,
the probability mass function
of the observables.

As illustrated by this example, auxiliary variables 
in statistical models 
are often introduced in order to obtain nice state-space models. 

In traditional statistical models, such auxiliary state variables
are themselves random variables, and the total model is a joint probability law
over all variables as, e.g., in (\ref{eqn:HMM}). 
(A statistical model may also contain parameters in addition 
to auxiliary random variables, 
but such parameters are not relevant for the present discussion.) 

The first main point of this paper is this:
\emph{requiring the auxiliary state variables to be random variables 
may be unnecessarily restrictive.}
The benefits of state-space representations
may be obtained by merely requiring 
a function $f(y,x)$ (with a useful factorization)
such that the probability mass function of the observables is
\begin{equation}
p(y) = \sum_{x} f(y,x);
\end{equation}
the function $f(y,x)$ need not be a probability mass function 
and it need not even be real valued.

\begin{figure}
\begin{center}
\begin{picture}(50,42)(0,-1)
\put(0,25){\dashbox(50,16){}}        \put(-1.5,32.5){\pos{cr}{$g$}}
\put(5,30){\framebox(5,5){}}         \put(7.5,36.25){\pos{cb}{$g_1$}}
\put(10,32.5){\line(1,0){12.5}}      \put(16.25,33.5){\pos{cb}{$X_1$}}
\put(22.5,30){\framebox(5,5){}}      \put(25,36.25){\pos{cb}{$g_2$}}
\put(27.5,32.5){\line(1,0){12.5}}    \put(33.75,33.5){\pos{cb}{$X_2$}}
\put(40,30){\framebox(5,5){}}        \put(42.5,36.25){\pos{cb}{$g_3$}}
\put(7.5,30){\line(0,-1){20}}        \put(8.5,20){\pos{cl}{$Y_1$}}
\put(25,30){\line(0,-1){20}}         \put(26,20){\pos{cl}{$Y_2$}}
\put(42.5,30){\line(0,-1){20}}       \put(43.5,20){\pos{cl}{$Y_3$}}
\put(0,-1){\dashbox(50,16){}}         \put(-1.5,7.5){\pos{cr}{$\ccj{g}$}}
\put(5,5){\framebox(5,5){}}          \put(7.5,4){\pos{ct}{$\ccj{g}_1$}}
\put(10,7.5){\line(1,0){12.5}}       \put(16.25,8.5){\pos{cb}{$X_1'$}}
\put(22.5,5){\framebox(5,5){}}       \put(25,4){\pos{ct}{$\ccj{g}_2$}}
\put(27.5,7.5){\line(1,0){12.5}}     \put(33.75,8.5){\pos{cb}{$X_2'$}}
\put(40,5){\framebox(5,5){}}         \put(42.5,4){\pos{ct}{$\ccj{g}_3$}}

\end{picture}
\caption{\label{fig:BasicExampleQFG}%
Factor graph for $p(y_1,y_2,y_3)$ with complex-valued factors.
}
\end{center}
\end{figure}
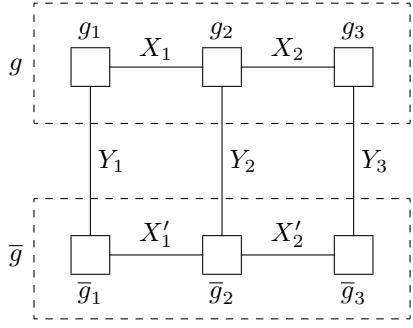

For example, 
consider the factor graph in \Fig{fig:BasicExampleQFG},
where all factors are complex valued. 
Note that the lower dashed box in \Fig{fig:BasicExampleQFG}
mirrors the upper dashed box, but all factors in the lower box
are the complex conjugates of the corresponding factors in the upper dashed box. 
The closed-box function of the upper dashed box is 
\begin{equation}
g(y_1,y_2,y_3) \eqdef \sum_{x_1,x_2} g_1(x_1,y_1) g_2(x_1,x_2,y_2) g_3(x_2,y_3)
\end{equation}
and the closed-box function of the lower dashed box is
\begin{equation}
\sum_{x_1',x_2'} \ccj{g_1(x_1',y_1)}\, \ccj{g_2(x_1',x_2',y_2)}\, \ccj{g_3(x_2',y_3)}
=  \ccj{g(y_1,y_2,y_3)}.
\end{equation}
If follows that the closed-box function in \Fig{fig:BasicExampleQFG}
(with both dashed boxes closed) is 
\begin{equation}
g(y_1,y_2,y_3) \ccj{g(y_1,y_2,y_3)} = |g(y_1,y_2,y_3)|^2,
\end{equation}
which is real and nonnegative 
and thus suitable to represent a probability mass function $p(y_1,y_2,y_3)$ (up to a scale factor).

We will see that factor graphs as in \Fig{fig:BasicExampleQFG}%
---with two parts, one part being the complex conjugate mirror image of the other part---%
can represent probabilities in quantum mechanics.

\section{Factor Graphs for Quantum Mechanics}
\label{sec:QMFG}

Consider the factor graph of \Fig{fig:GenQFG}. 
In this figure, $U_0$ and $U_1$ are $M\times M$ unitary matrices, 
and all variables except $Y_1$ and $Y_2$ take values in the set $\{ 1,\ldots, M\}$. 
The two large boxes in the figure represent measurements, as will be detailed below. 
The factor/box $p(x_0)$ is a probability mass function over the initial state $X_0$.
We will see that this factor graph 
(with suitable modeling of the measurements) 
represents the joint probability mass function $p(y_1,y_2)$
of a general $M$-dimensional quantum system with two observations $Y_1$ and $Y_2$.
The generalization to more observed variables $Y_1, Y_2, \ldots$ is obvious.

The unitary matrices $U_0$ and $U_1$ in \Fig{fig:GenQFG} represent the development 
of the system between the initial state and the first measurement, 
or between measurements, respectively, according to the Schr\"odinger equation.

In the most basic case, 
the initial state $X_0=x_0$ is known 
and the measurements look as shown in \Fig{fig:BasicQMFG},
where the matrices $B_1$ and $B_2$ are also unitary.
In this case, the observed variables $Y_1$ and $Y_2$ 
take values in $\{ 1,\ldots, M\}$ as well. 
Note that the lower part of this factor graph is the complex conjugate mirror of the upper part 
(as in \Fig{fig:BasicExampleQFG}).

In quantum-mechanical terms, measurements 
as in \Fig{fig:BasicQMFG} are projection measurements with one-dimensional eigenspaces. 
Note that the value of $Y_1$ and $Y_2$ is the index of the measured eigenspace 
(rather than the corresponding eigenvalue).

\begin{figure}
\setlength{\unitlength}{0.85mm}
\begin{center}
\begin{picture}(40,51)(0,-10)
\put(0,32.5){\line(1,0){15}}        \put(4,33.5){\pos{cb}{$X_k$}}
\put(15,30){\framebox(5,5){}}       \put(17.5,36.5){\pos{cb}{$A_k$}}
 \put(20,32.5){\markerDot}
\put(20,32.5){\line(1,0){20}}       \put(36,33.5){\pos{cb}{$\tilde{X}_k$}}
\put(17.5,30){\line(0,-1){7.5}}     
\put(15,17.5){\framebox(5,5){$=$}}
 \put(20,20){\line(1,0){5}}
 \put(25,20){\line(0,-1){30}}      \put(24,-10){\pos{br}{$Y_k$}}
\put(17.5,17.5){\line(0,-1){7.5}}
\put(0,7.5){\line(1,0){15}}         \put(4,8.5){\pos{cb}{$X_k'$}}
\put(15,5){\framebox(5,5){}}        \put(17.5,3.5){\pos{ct}{$A_k^\H$}}
 \put(15,7.5){\markerDot}
\put(20,7.5){\line(1,0){20}}        \put(36,8.5){\pos{cb}{$\tilde{X}_k'$}}
\put(10,-2){\dashbox(20,43){}}
\end{picture}
\caption{\label{fig:GenMeasurement}%
General measurement as in \cite[Chap.~2]{NiChuang:QCI}. 
Condition (\ref{eqn:GenMeasurementCond}) must be satisfied.
}
\end{center}
\end{figure}
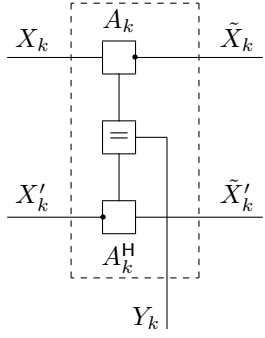

A very general form of measurement is shown in \Fig{fig:GenMeasurement}. 
In this case, the range of $Y_k$ is a finite set $\calY_k$, 
and for each $y\in \calY_k$, 
the factor $A_k(\tilde{x}_k,x_k,y)$ corresponds to a complex square matrix $A_k(y)$ 
(with row index $\tilde{x}_k$ and column index $x_k$) such that 
\begin{equation} \label{eqn:GenMeasurementCond}
\sum_{y\in J_k} A_k(y)^\H A_k(y) = I
\end{equation}
(see \cite[Chap.~2]{NiChuang:QCI}). 
Measurements as in \Fig{fig:BasicQMFG}
are included as a special case with $\calY_k = \{1,\ldots, M\}$ and 
\begin{equation}
A_k(y) = B_k(y) B_k(y)^\H,
\end{equation}
where $B_k(y)$ denotes the $y$-th column of $B_k$.

\begin{figure*}[p]
\centering
\begin{picture}(117.5,40)(-5,-10)
\put(0,7.5){\framebox(5,5){}}      \put(2.5,6){\pos{ct}{$p(x_0)$}}
\put(5,10){\line(1,0){12.5}}       \put(11.25,11){\pos{cb}{$X_0$}}
\put(20,17.5){\line(1,0){10}}
 \put(20,17.5){\line(0,-1){5}}
\put(17.5,7.5){\framebox(5,5){$=$}}
 \put(20,7.5){\line(0,-1){5}}
\put(20,2.5){\line(1,0){10}}
\put(30,15){\framebox(5,5){}}     \put(32.5,13.5){\pos{ct}{$U_0$}}
 \put(35,17.5){\markerDot}
\put(35,17.5){\line(1,0){12.5}}   \put(41.25,18.5){\pos{cb}{$X_1$}}
\put(30,0){\framebox(5,5){}}      \put(32.5,-1.5){\pos{ct}{$U_0^\H$}}
 \put(30,2.5){\markerDot}
\put(35,2.5){\line(1,0){12.5}}    \put(41.25,3.5){\pos{cb}{$X_1'$}}
\put(47.5,0){\framebox(10,20){}}
\put(52.5,0){\line(0,-1){7.5}}    
 \put(52.5,-7.5){\knownBox}       \put(51.5,-6){\pos{br}{$Y_1=y_1$}}
\put(57.5,17.5){\line(1,0){12.5}} \put(64.5,18.5){\pos{cb}{$\tilde{X}_1$}}
\put(70,15){\framebox(5,5){}}     \put(72.5,13.5){\pos{ct}{$U_1$}}
 \put(75,17.5){\markerDot}
\put(75,17.5){\line(1,0){12.5}}   \put(81.25,18.5){\pos{cb}{$X_2$}}
\put(57.5,2.5){\line(1,0){12.5}}  \put(64.5,3.5){\pos{cb}{$\tilde{X}_1'$}}
\put(70,0){\framebox(5,5){}}      \put(72.5,-1.5){\pos{ct}{$U_1^\H$}}
 \put(70,2.5){\markerDot}
\put(75,2.5){\line(1,0){12.5}}    \put(81.25,3.5){\pos{cb}{$X_2'$}}
\put(87.5,0){\framebox(10,20){}}
\put(92.5,0){\line(0,-1){7.5}}    \put(93.5,-7.5){\pos{bl}{$Y_2$}}
\put(97.5,17.5){\line(1,0){12.5}}  \put(103,18.5){\pos{cb}{$\tilde{X}_2$}}
\put(110,17.5){\line(0,-1){5}}
\put(107.5,7.5){\framebox(5,5){$=$}}
\put(110,7.5){\line(0,-1){5}}
\put(97.5,2.5){\line(1,0){12.5}}   \put(103,3.5){\pos{cb}{$\tilde{X}_2'$}}
\put(-5,-10){\dashbox(65.5,35){}}    \put(27.75,26.5){\pos{cb}{$\propto\tilde{\rho}_1$}}
\end{picture}
\caption{\label{fig:PostMeasurementDensityMatrix}%
Density matrix $\tilde{\rho}_1$ after measuring $Y_1=y_1$.
}
\end{figure*}
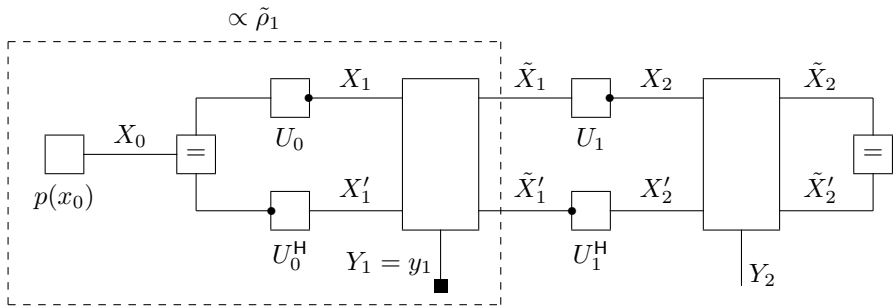

It is clear from Section~\ref{sec:StatModel} that 
the external function of \Fig{fig:GenQFG} 
(with measurements as in \Fig{fig:BasicQMFG} or as in \Fig{fig:GenMeasurement})
is real and nonnegative. 
We now proceed to analyze these factor graphs and to verify that
they yield the correct quantum-mechanical probabilities $p(y_1,y_2)$
for the respective class of measurements.
To this end, 
we need to understand the closed-box functions 
of the dashed boxes in \Fig{fig:BackwardMesg}.
We begin with the dashed box on the right-hand side of \Fig{fig:BackwardMesg},
where $Y_2$ is assumed to be unknown. 

\begin{proposition} \label{prop:QMBackwardBoxes}
Closing the dashed box on the right-hand side in \Fig{fig:BackwardMesg}
(with a measurement of $Y_2$ as in \Fig{fig:BasicQMFG} or as in \Fig{fig:GenMeasurement},
but with unknown result of the measurement)
reduces it to an equality constraint function.
\end{proposition}
The proof of this proposition and the proofs of the subsequent propositions are easy, 
and are omitted due to space constraints.

Proposition~\ref{prop:QMBackwardBoxes} 
guarantees, in particular, that a future measurement (with as yet unknown results) 
does not influence present or past observations. 

The proposition clearly holds also for the extension of 
\Fig{fig:GenQFG} to any finite number of measurements $Y_1, Y_2, \ldots$ and can 
then be applied recursively from right to left. 

Applying reductions according to Proposition~\ref{prop:QMBackwardBoxes}
recursively from right to left in \Fig{fig:GenQFG} leads to the following proposition.

\begin{proposition} \label{prop:ProperlyNormalized}
The factor graph of \Fig{fig:GenQFG}
(with measurements as in \Fig{fig:BasicQMFG} or as in \Fig{fig:GenMeasurement})
represents a properly normalized probability mass function, i.e., 
the external function $p(y_1,y_2)$ is real and $\sum_{y_1,y_2} p(y_1,y_2) = 1$.
\end{proposition}

Consider now the dashed box on the left in Figs.\ \ref{fig:BackwardMesg} 
and~\ref{fig:PostMeasurementDensityMatrix}, 
which turns out to be the density matrix of quantum mechanics. 
We will distinguish between the closed-box function 
$\rho_k(x_k,x_k')$ before measuring $Y_k$ (as in \Fig{fig:BackwardMesg}), 
and the closed-box function 
$\tilde{\rho}_k(\tilde{x}_k,\tilde{x}_k')$ 
after the observation $Y_k=y_k$ (as in \Fig{fig:PostMeasurementDensityMatrix}). 
The former is easily seen to be properly normalized, but the latter needs normalization 
to satisfy (\ref{eqn:normalizedRho}). 
The corresponding matrices will be denoted by $\rho_k$ and $\tilde{\rho}_k$, respectively.
The proper normalization can then be expressed by the condition
\begin{equation} \label{eqn:normalizedRho}
\tr(\rho_k) = \tr(\tilde{\rho}_k) = 1.
\end{equation}

\begin{proposition}[Unitary Evolution Between Measurements]
The matrix $\rho_{k+1}$ is obtained from the matrix $\tilde{\rho}_k$ as
\begin{equation}
\rho_{k+1} = U_k \tilde{\rho}_k U_k^\H.
\end{equation}
\eproofnegspace
\end{proposition}

\begin{proposition}[Basic Projection Measurement]
In \Fig{fig:GenQFG} (generalized to any number of observations), 
if $Y_k$ is measured as in \Fig{fig:BasicQMFG}, then
\begin{IEEEeqnarray}{rCl}
\IEEEeqnarraymulticol{3}{l}{
P(Y_k = y \mid Y_{k-1}=y_{k-1},\ldots,Y_1=y_1) 
}\nonumber\\ \quad
 & = &  B_k(y)^\H \rho_k B_k(y)  \label{eqn:BasicProjectionProbability1} \\
 & = &  \tr\!\left( B_k(y) B_k(y)^\H \rho_k \right),
\end{IEEEeqnarray}
where $B_k(y)$ is the $y$-th column of $B_k$.
After measuring/observing $Y_k=y$, 
the density matrix is 
\begin{equation} \label{eqn:BasicProjectionPostMeasurementRho}
\tilde{\rho}_k = B_k(y) B_k(y)^\H.
\end{equation}
\eproofnegspace
\end{proposition}
Note that (\ref{eqn:BasicProjectionPostMeasurementRho}) 
is properly normalized because 
$\tr(B_k(y) B_k(y)^\H) = \tr(B_k(y)^\H B_k(y)) = \|B_k(y)\|^2 = 1$.

In the special case of \Fig{fig:BasicQMFG}, 
with known initial state \mbox{$X_0=x_0$}, 
the matrix $\rho_k$ factors as 
\begin{equation}
\rho_k(x_k,x_k') = \psi_k(x_k) \ccj{\psi_k(x_k')},
\end{equation}
or, in matrix notation,
\begin{equation}
\rho_k = \psi_k \psi_k^\H,
\end{equation}
where $\psi_k$ is a column vector of norm~1. 
The post-measurement density matrix $\tilde{\rho}_k$ factors analoguously,
as is obvious from (\ref{eqn:BasicProjectionPostMeasurementRho}).
In quantum-mechanical terms, $\psi_k$ is the quantum state. 
The probability (\ref{eqn:BasicProjectionProbability1})
can then be expressed as 
\begin{IEEEeqnarray}{rCl}
\IEEEeqnarraymulticol{3}{l}{
P(Y_k = y \mid Y_{k-1}=y_{k-1},\ldots,Y_1=y_1) 
}\nonumber\\ \quad
 & = &  B_k(y)^\H \psi_k \psi_k^\H B_k(y) \\
 & = &  \| B_k(y)^\H \psi_k \|^2,
\end{IEEEeqnarray}
which is the most basic form of computing probabilities in quantum mechanics.

\begin{proposition}[General Measurement] \label{prop:GeneralMeasurement}
In \Fig{fig:GenQFG} (generalized to any number of observations), 
if $Y_k$ is measured as in \Fig{fig:GenMeasurement}, then
\begin{IEEEeqnarray}{rCl}
\IEEEeqnarraymulticol{3}{l}{
P(Y_k = y \mid Y_{k-1}=y_{k-1},\ldots,Y_1=y_1) 
}\nonumber\\ \quad
 & = &  \tr\!\left( A_k(y) \rho_k A_k(y)^\H \right).
\end{IEEEeqnarray}
After measuring/observing $Y_k=y$, the density matrix is 
\begin{equation}
\tilde{\rho}_k = \frac{A_k(y) \rho_k A_k(y)^\H}{\tr\!\left( A_k(y) \rho_k A_k(y)^\H \right)}
\end{equation}
\eproofnegspace
\end{proposition}

According to 
Propositions \ref{prop:ProperlyNormalized}--\ref{prop:GeneralMeasurement}, 
the factor graph of \Fig{fig:GenQFG} 
(with measurements as in \Fig{fig:BasicQMFG} or as in \Fig{fig:GenMeasurement})
yields indeed the correct quantum-mechanical probabilities for the respective class of measurements.

\section{Conclusion}
\label{sec:Concl}

We have proposed a class of factor graphs that represent 
quantum-mechanical probabilities involving any number of measurements,
both for basic projection measurements and for general measurements as in \cite[Chap.~2]{NiChuang:QCI}. 
Such factor graphs have not previously been used in statistical modeling. 

The space constraints of this paper preclude the discussion of further
pertinent topics that we intend to address elsewhere, including the meaning of
such factor graphs from a statistical-modeling point of view (disregarding
physics), the relation to quantum Bayesian networks (see, e.g., \cite{Tu:qi}),
to quantum belief propagation (see, e.g., \cite{LePo:qbp2008}), and to tensor
diagrams/networks for analyzing quantum systems (see, e.g.,
\cite{GuLeWe:prb2008}). It is also noteworthy that quantum circuits as in
\cite[Chap.~4]{NiChuang:QCI} may be viewed as halves of factor graphs as in
this paper, where the missing other half is a complex conjugate mirror image
as in Section~\ref{sec:StatModel}.

{\small

\newcommand{\IT}{IEEE Trans.\ Inf.\ Theory}
\newcommand{\COM}{IEEE Trans.\ Communications}
\newcommand{\ComLett}{IEEE Communications Letters}
\newcommand{\JSAC}{IEEE J.\ on Selected Areas in Communications}
\newcommand{\WirelessComm}{IEEE Trans.\ Wireless Communications}
\newcommand{\SP}{IEEE Trans.\ Signal Processing}
\newcommand{\SPMag}{IEEE Sig.\ Proc.\ Mag.}
\newcommand{\ProcIEEE}{Proceedings of the IEEE}

} 


\begin{thebibliography}{99}

\bibitem{KFL:fg2000}
  F.~R.\ Kschi\-schang, B.~J.\ Frey, and H.-A.\ Loeliger,
  ``Factor graphs and the sum-product algorithm,''
  \emph{\IT,} vol.~47, pp.~498--519, Feb.\ 2001.

\bibitem{Lg:ifg2004}
H.-A. Loeliger,
``An introduction to factor graphs,''
\emph{\SPMag,} Jan.\ 2004, pp.~28--41.

\bibitem{LDHKLK:fgsp2007}
H.-A.~Loeliger, J.~Dauwels, Junli~Hu, S.~Korl, Li~Ping, and F.~R.~Kschi\-schang,
``The factor graph approach to model-based signal processing,''
\emph{\ProcIEEE,} vol.~95, no.~6, pp.~1295--1322, June 2007.


\bibitem{MeMo:ipc}
M.~M\'ezard and A.~Montanari,
\emph{Information, Physics, and Computation.}
Oxford University Press, 2009.


\bibitem{Jo:gm2004}
M.~I.~Jordan,
``Graphical models,''
\emph{Statistical Science,} vol.~19, no.~1, pp.~140--155, 2004.

\bibitem{bi:prml}
Ch.~M.~Bishop,
\emph{Pattern Recognition and Machine Learning.}
New York: Springer Science+Business Media, 2006.


\bibitem{KoFr:PGMb}
D.~Koller and N.~Friedman,
\emph{Probabilistic Graphical Models.}
Cambridge, MA, MIT Press, 2009.


\bibitem{AFP:QM}
G.~Auletta, M.~Fortunato, and G.~Parisi,
\emph{Quantum Mechanics.}
Cambridge University Press, 2009.

\bibitem{NiChuang:QCI}
M.~A.~Nielsen and I.~L.~Chuang,
\emph{Quantum Computation and Quantum Information.}
Cambridge University Press, 2000.


\bibitem{Cv:gtbt}
P.~Cvitanovi{\'c},
\emph{Group Theory: Birdtracks, Lie's, and Exceptional Groups.}
Princeton Univ.\ Press, 2008.

\bibitem{Pe:ula2009}
E.~Peterson,
``Unshackling linear algebra from linear notation,''
arXiv:0910.1362, 2009.

\bibitem{BaMa:nfght2011}
A.~Al-Bashabsheh and Y.~Mao,
``Normal factor graphs and holographic transformations,''
\emph{\IT,} vol.~57, no.~2, pp.~752--763, Feb.\ 2011.

\bibitem{FoVo:pfnfg2011c}
G.~D.\ Forney, Jr., and P.~O.\ Vontobel,
``Partition functions of normal factor graphs,''
\emph{Proc.\ Inf.\ Theory \& Appl.\ Workshop,}
UC San Diego, La Jolla, CA, USA, Feb.~6--11, 2011.

\bibitem{BMV:nfgla2011c}
A.~Al-Bashabsheh, Y.~Mao, and P.~O.~Vontobel,
``Normal factor graphs: a diagrammatic approach to linear algebra,''
\emph{Proc.\ IEEE Int.\ Symp.\ Inf.\ Theory,} 
St.~Petersburg, Russia, Jul.~31--Aug.~5, 2011, pp.~2178--2182.


\bibitem{Tu:qi}
R.~R.~Tucci, 
``Quantum information theory -- a quantum Bayesian net perspective,''
arXiv:quant-ph/9909039v1, 1999.


\bibitem{LePo:qbp2008}
M.~S.\ Leifer and D.~Poulin,
``Quantum graphical models and belief propagation,''
\emph{Annals of Physics,} vol.~323, no.~8, pp.~1899--1946, Aug.\ 2008.

\bibitem{GuLeWe:prb2008}
Z.-C.\ Gu, M.\ Levin, and X.-G.\ Wen,
``Tensor-entanglement renormalization group approach as a unified method for
symmetry breaking and topological phase transitions,''
\emph{Phys.\ Rev.\ B,} vol.~78, p.~205116, Nov.\ 2008.





\end{thebibliography}
\end{document}